\documentclass[english, 11pt, a4paper]{article}

\usepackage{slashed}
\usepackage[utf8]{inputenc}
\usepackage{babel}
\usepackage{graphics}
\usepackage{graphicx}
\usepackage{texdraw}
\usepackage{epsfig}
\usepackage{float}
\usepackage{amsmath}
\usepackage{amssymb}
\usepackage{enumerate}
\usepackage[all]{xy}
\usepackage{fullpage}
\usepackage[small]{caption}
\usepackage{hyperref}

\newcommand{\be}{\begin{equation}}
\newcommand{\ee}{\end{equation}}

\begin{document}

\title{
Topology in the $SU(N_f)$ chiral symmetry restored phase of unquenched QCD and            
  axion cosmology
}

\author{Vicente ~Azcoiti\footnote{Acknowledges financial support by
    Ministerio de Economía y Competitividad under Grant No. FPA2015-65745-P (MINECO/FEDER).}\\
    Departamento de F{\'i}sica Te\'orica, Facultad de Ciencias \\
    Universidad de Zaragoza, Pedro Cerbuna 9, 50009 Zaragoza, Spain\\}

\maketitle

\begin{abstract}

  The axion is one of the more interesting candidates to make the dark matter of the universe,
  and the axion potential plays a fundamental role in the determination of the dynamics of the
  axion field. Moreover, the way in which the $U(1)_A$ anomaly manifests itself in the chiral
  symmetry restored phase of $QCD$ at high temperature could be tested when probing the
  $QCD$ phase transition in relativistic heavy ion collisions. With these motivations, we
  investigate the physical consequences of the survival of the effects of the $U(1)_A$ anomaly
  in the chiral symmetric phase of $QCD$, and show that the free energy density is a singular
  function of the quark mass $m$, in the chiral limit, and that the $\sigma$ and $\bar\pi$
  susceptibilities diverge in this limit at any $T\ge T_c$. We also show that the difference
  between the $\bar\pi$ and $\bar\delta$ susceptibilities diverges in the chiral limit at any
  $T\ge T_c$, a result that can be contrasted with the existing lattice calculations; and
  discuss on the generalization of these results to the $N_f\ge 3$ model.

\end{abstract}

\vfill\eject

\section{Introduction}\label{intro}

We report the results of an investigation on the physical consequences of the survival of
the topological effects of the axial anomaly in the high temperature phase of $QCD$. This
contribution to the Lattice 2017 Symposium is based on references \cite{trece}, \cite{vic2}
and we refer the interested reader to these papers.
To summarize the main results, our starting hypothesis in \cite{trece} was to assume that
the perturbative expansion of the free energy density in powers of the quark mass, $m$, has a
non-vanishing convergence radius in the high temperature chiral symmetric phase of $QCD$. This
is just what we expect on physical grounds if all correlation lengths remain finite in the
chiral limit, and the spectrum of the model shows therefore a mass gap also in this limit.
The main conclusion obtained from this hypothesis was that all the topological effects
of the axial anomaly should disappear in this phase, the topological susceptibility and all
$\theta$-derivatives of the free energy density vanish, and the theory becomes
$\theta$-independent at any $T > Tc$ in the infinite-volume limit. Accordingly, the free
energy density should be a singular function of the quark mass, in the chiral limit, if the
topological effects of the $U(1)_A$ anomaly survive in the chiral symmetry restored phase of
$QCD$ at finite temperature, and the main purpose of reference \cite{vic2} was to
investigate this issue. The starting hypothesis in \cite{vic2} was to assume that the
topological effects of the anomaly survive in the high temperature phase of QCD, and the
model shows therefore a non-trivial $\theta$-dependence in this phase. Under this
assumption we showed that indeed the free energy density is a singular function of the quark
mass, $m$, in the chiral limit, at any $T > T_c$, and that the correlation length and the
$\sigma$ and $\bar\pi$ susceptibilities diverge in this limit, as well as the difference
between the $\bar\pi$ and $\bar\delta$ susceptibilities.

The relevance of these results is due to the fact that the topological effects of the $U(1)_A$
axial anomaly in the high temperature phase of $QCD$ play a fundamental role in
the determination of the dynamics of the axion field, which is one of the
more interesting candidates to make the dark matter of the universe. Moreover these results
could be tested when probing the $QCD$ phase transition in relativistic heavy ion collisions.

The first investigations on this subject started long time ago.
The idea that the chirally restored phase of two-flavor $QCD$ is symmetric under
$U(2)\times U(2)$
rather than $SU(2)\times SU(2)$ was raised by Shuryak in 1994 \cite{shu} based on an
instanton liquid-model study.
In 1996 Cohen \cite{cohen1} also got this result formally from the QCD functional integral
under some assumptions.
However immediately after several calculations questioning this result appeared
\cite{cuatro}-\cite{boy}. Since then much work has been done, and references
\cite{trece}-\cite{vic2}, \cite{diez}-\cite{javier} contain recent
work concerning theoretical developments, the computation of the topological susceptibility
(with discrepant results, as shown in this conference), and the computation of correlation
functions, spectral density of the Dirac operator and susceptibilities.

\section{$\sigma$ and $\eta$ susceptibilities}\label{sec-1}

Our starting point is the Euclidean continuum Lagrangian of $N_f$ flavors $QCD$ with a
$\theta$-term

\begin{equation}
  L = \sum_f L^f_F + \frac{1}{4} F^a_{\mu\nu}\left(x\right)F^a_{\mu\nu}\left(x\right)
  + i\theta \frac{g^2}{64\pi^2} \epsilon_{\mu\nu\rho\sigma}
  F^a_{\mu\nu}\left(x\right)F^a_{\rho\sigma}\left(x\right)
  \label{eulagran}
\end{equation}
where $L^f_F$ is the fermion Lagrangian for the $f$-flavor, and

\begin{equation}
  Q = \frac{g^2}{64\pi^2} \int d^4x\hskip 0.1cm \epsilon_{\mu\nu\rho\sigma}
  F^a_{\mu\nu}\left(x\right)F^a_{\rho\sigma}\left(x\right)
  \label{ftopcharg}
\end{equation}
is the topological charge of the gauge configuration.

To avoid ultra-violet divergences we will assume Ginsparg-Wilson fermions, which show an
$U(1)_A$ anomalous symmetry, good chiral properties, a quantized topological charge
and an exact index theorem on the lattice, so they share all essential ingredients with the
continuum formulation. We will also assume in what follows that the topological effects of
the $U(1)_A$ axial anomaly survive in the high temperature chiral symmetric phase of QCD, and
that the partition function and the free energy density

\begin{equation}
  Z\left(\theta\right) = e^{-V_x L_t E\left(\beta,m,\theta\right)}
  \label{zetalarge}
\end{equation}
show a non-trivial dependence on the $\theta$ parameter.
$E\left(\beta,m,\theta\right)$ in (\ref{zetalarge}) is the free energy density, $\beta$ the
inverse gauge coupling, $m$ the quark mass, and $L_t$ the lattice temporal extent or inverse
physical temperature $T$. Moreover the mean
value of any intensive operator $O$, as for instance the scalar and pseudoscalar
condensates, or any correlation function, in the $Q=0$ topological sector, can be
computed as

\begin{equation}
  \left< O\right>_{Q=0} = \frac{\int d\theta \left< O\right>_\theta Z(\theta, m)}
         {\int d\theta Z(\theta,m)}
         \label{mascurioso}
\end{equation}
with $\left< O\right>_\theta$ the mean value of $O$ computed with the integration
measure (\ref{eulagran}). Because the free energy density, as a
function of $\theta$, has its absolute minimum at $\theta=0$ for non-vanishing quark
masses, the following relation holds in the infinite lattice volume limit

\begin{equation}
  \left< O\right>_{Q=0} = \left< O\right>_{\theta=0}
  \label{mascuriosob}
\end{equation}
and we want to remark that, as discussed in \cite{trece}, in spite of the fact that the
$Q=0$ topological sector is free from the global $U(1)_A$ anomaly, and spontaneously breaks
the $U(N_f)_A$ axial symmetry at $T=0$, equation (\ref{mascuriosob}) is
compatible with a massive flavor-singlet pseudoscalar meson in the chiral limit.

Let us consider, for simplicity, the two-flavor model with degenerate up and down
quark masses. In the high temperature phase
the $SU(2)_A$ symmetry is fulfilled in the ground state for massless quarks, and therefore
the mean value of the flavor singlet scalar condensate $\left< S\right>$, as well as
of any order parameter
for this symmetry, vanishes in the chiral limit.
Moreover the infinite lattice volume limit and the chiral limit should commute,
provided the order parameter remains bounded. In addition equation
(\ref{mascuriosob}) implies that the $SU(2)_A$ symmetry is also fulfilled in the
the $Q=0$ topological sector. However, the $U(1)_A$ symmetry should be spontaneously
broken in this sector, giving account in this way for the $U(1)_A$ anomaly\footnote{The
  Goldstone theorem however can be fulfilled without a Nambu-Goldstone boson \cite{trece}.}.

Let us assume that the flavor singlet scalar susceptibility, the $\sigma$-susceptibility,
remains finite when we approach the chiral limit. Then the flavor singlet scalar condensate
at $\theta=0$, which is equal to the mean value of the same operator in the $Q=0$ sector,
shows this linear behavior

\begin{equation}
\left< S\right>_{\theta=0} =
\left< S\right>_{{Q=0}_{\hskip 0.1cmm\rightarrow 0}} \approx \chi_{\sigma}\left(0\right) m
\label{linear}
\end{equation}
with the quark mass for small quark masses. Because the $Q=0$ sector is free from the
global $U(1)_A$ anomaly, we can write the following Ward identities

\begin{equation}
\chi_{\bar\pi}\left(m\right)_{Q=0} =
\chi_\eta\left(m\right)_{Q=0} = \frac{\left\langle S\right\rangle_
  {Q=0}}{m}\rightarrow \chi_{\sigma}\left(0\right)
\label{ward}
\end{equation}
which tell us that the pion, eta and sigma susceptibilities in the
chiral limit in this sector are equal to the sigma susceptibility in full QCD at $\theta=0$.
Moreover we can demonstrate, with the help of an anomalous $U(1)_A$ transformation, the
following equation

\begin{equation}
\left\langle S\left(x\right)S\left(0\right)\right\rangle^{m=0}_{Q=0} =
\frac{1}{2}
\left\langle S\left(x\right)S\left(0\right)\right\rangle^{m=0}_{\theta=0} +
\frac{1}{2}
\left\langle P\left(x\right)P\left(0\right)\right\rangle^{m=0}_{\theta=0}
\label{susrelation}
\end{equation}
which relates the flavor singlet scalar correlation function in the $Q=0$ sector in
the chiral limit, with the same quantity, and the eta-correlation function
in full $QCD$.

Equation (\ref{susrelation}) implies the following relation

\begin{equation}
\chi_{\sigma}\left(0\right) =
\chi_{\sigma}\left(0\right)_{Q=0} = \frac{
  \chi_{\sigma}\left(0\right) + \chi_{\eta}\left(0\right)}{2}
\label{absurdo}
\end{equation}
between the $\sigma$ and $\eta$ susceptibilities in full $QCD$.
The fulfillment of this equation requires the equality of the sigma and eta susceptibilities
in the chiral limit, in contradiction with the assumption that the topological effects of
the $U(1)_A$ axial anomaly survive in the high temperature phase of QCD. We conclude
therefore that the assumption on the finitude of the correlation length and
$\sigma$-susceptibility in the chiral limit is not compatible with the survival of
the topological effects of the
$U(1)_A$ anomaly in the high temperature phase of $QCD$. Hence the
$\sigma$-susceptibility should diverge in the chiral limit, and the free
energy density should be singular at vanishing quark masses.

Following the standard wisdom, the critical behavior of the model should be well described by
a power law behavior for the flavor singlet scalar condensate

\begin{equation}
\left< S\right>_{{\theta=0}_{\hskip 0.1cmm\rightarrow 0}} \approx C\left(T\right)
m^{\frac{1}{\delta}}
\label{powerlaw}
\end{equation}
with $\delta>1$. But equation (\ref{powerlaw}) implies that the flavor singlet scalar
susceptibility diverges at any $T\ge T_c$ in the chiral limit

\begin{equation}
\chi_\sigma\left(m\right) \approx C\left(T\right) \frac{1}{\delta}
m^{\frac{1-\delta}{\delta}}
\label{diverges}
\end{equation}
and because the $SU(2)_A$
symmetry is not anomalous, the pion susceptibility verifies this Ward identity

\begin{equation}
  \chi_{\bar\pi}\left(m\right) = \frac{\left\langle S\right\rangle}{m}
  \label{ward2}
\end{equation}
and also diverges in the chiral limit this way

\begin{equation}
\chi_{\bar\pi}\left(m\right) \approx C\left(T\right)
m^{\frac{1-\delta}{\delta}}.
\label{diverges2}
\end{equation}
Moreover the vector meson $\bar\delta$ susceptibility, $\chi_{\bar\delta}$, which is bounded
by the scalar susceptibility, $\chi_\sigma$ , verifies the following inequality

\begin{equation}
\chi_{\bar\pi}\left(m\right) - \chi_{\bar\delta}\left(m\right) \ge
\chi_{\bar\pi}\left(m\right) - \chi_\sigma\left(m\right) \approx C\left(T\right)
\frac{\delta-1}{\delta}
m^{\frac{1-\delta}{\delta}}
\label{diverges3}
\end{equation}
which shows that this quantity, which is an order parameter
for the $U(1)_A$ axial symmetry, also diverges in the chiral limit.

The zero-temperature two-flavor Schwinger model is a good toy model for testing these results.
The $SU(2)_A$
chiral symmetry is fulfilled in the vacuum in this model because of the Coleman-Mermin-Wagner
theorem. Moreover it is well known that the topological effects of the $U(1)_A$ anomaly are
relevant in this model since it shows a non-trivial dependence on the
$\theta$-parameter \cite{coleman}. It has been shown that the free energy density
is singular in the chiral limit \cite{smilgasch} in this model, and that the pions are
massless in this limit \cite{gatt}; in agreement with the discussion developed in this section.

\section{Phase diagram of QCD in the Q=0 topological sector}\label{sec-2}

These results can also be shown by performing a qualitative analysis of the phase diagram
of $QCD$ in the $Q=0$ sector. The $SU(2)_A$ symmetry is fulfilled in $QCD$
at any $T> T_c$, and therefore the up and down scalar condensates $\left< S_u\right>$,
$\left< S_d\right>$ vanish in the chiral limit $m_u = m_d = 0$. However if we
consider $QCD$ with two non degenerate quark flavors, and take the limit
$m_u\rightarrow 0$ keeping $m_d$ fixed, or vice versa, the condensate
$\left< S_u\right>$, or $\left< S_d\right>$, takes a non-vanishing mean
value due to the fact that the $U(1)_u$ symmetry at $m_u=0$, or the $U(1)_d$ symmetry
at $m_d=0$, which would enforce the condensate to be zero, is anomalous.
But since equation (\ref{mascuriosob}) can be applied to these condensates,
this result tell us
that the $Q=0$ topological sector, which is free from the global axial anomaly, spontaneously
breaks the $U(1)_u$ axial symmetry at $m_u=0, m_d\ne 0$, and the $U(1)_d$ symmetry at
$m_d=0, m_u\ne 0$. The phase diagram of $QCD$ in the $Q=0$ topological sector, in the
$(m_u, m_d)$ plane, shows therefore two first order phase transition lines, which coincide
with the coordinate axes, finishing at the end point $m_u=m_d=0$, which is a critical
point for any $T> T_c$.

Equation (\ref{mascuriosob}) tell us that the critical chiral equation of state of $QCD$
at $\theta=0$ should be the same as the one of the $Q=0$ topological sector, and
should show therefore a divergent correlation length at any $T>T_c$ in the chiral limit. We
expect therefore a
continuous finite temperature chiral transition, and a divergent correlation length
for any $T\ge T_c$, and
because the symmetry breaking pattern is, in the two flavor model,
$SU(2)_L\times SU(2)_R\rightarrow SU(2)_V$, the critical equation of state should
be that of the three-dimensional $O(4)$ vector universality class \cite{piswil},
which shows a critical exponent $\delta = 4.789(6)$ \cite{hasen} ($\delta = 3$
in the mean field or Landau approach).

For $N_f\ge 3$ a similar argument on the phase diagram of the $Q=0$ sector applies, but the
scenario that emerges in this case is not plausible because no stable fixed points
are expected in the corresponding Landau-Ginzburg-Wilson $\Phi^4$ theory compatible
with the given symmetry-breaking pattern \cite{ettore}.

\section{Conclusions and comments}\label{conclusions}

We started recently an investigation of the topological
properties of $QCD$ in the high temperature chiral symmetric phase in
reference \cite{trece}.
The starting hypothesis in \cite{trece} was to assume
that the perturbative expansion of the free energy density in powers of the quark
mass, $m$, has a non-vanishing convergence radius in the high temperature chiral
symmetric phase of $QCD$, which is just what we expect if all correlation lengths remain
finite in the chiral limit, and the spectrum of the model shows therefore a mass gap
also in this limit. The main conclusion in \cite{trece} was
that all the topological effects of the axial anomaly should disappear in this phase,
the topological susceptibility and all $\theta$-derivatives of the free energy
density vanish, and the theory becomes $\theta$ independent at any
$T > T_c$ in the infinite-volume limit. Accordingly, the free energy density should be a
singular function of the quark mass, in the chiral limit, if the topological effects
of the $U(1)_A$ anomaly survive in the chiral symmetry restored phase of QCD at finite
temperature.

Ongoing with this research line, the main purpose in reference \cite{vic2} was to further
investigate this issue. To this end our starting hypothesis was to assume
that the topological effects of the anomaly survive in the high temperature phase of
$QCD$, and the model shows therefore a
non-trivial $\theta$-dependence in this phase. Under this assumption we have shown
that indeed, the free energy density is a singular function of the quark mass, $m$,
in the chiral limit at any $T>T_c$, and that the correlation length and the
$\sigma$ and $\bar\pi$ susceptibilities diverge in this limit.
Under the same assumption we have also shown that the difference between the
$\bar\pi$ and $\bar\delta$ susceptibilities diverges in the chiral limit at any
$T\ge T_c$.

This result seems to be excluded by recent results of
Tomiya et al. \cite{catorce} from numerical simulations of two-flavor QCD, thus
suggesting the effects of the $U(1)_A$ anomaly are absent in the chiral symmetric phase
of two-flavor $QCD$.
However, previous results by Dick et al. \cite{sharma} on larger lattices, but using
overlap fermions only in the valence sector, seem to predict a divergent
$\chi_{\bar\pi}\left(m\right) - \chi_{\bar\delta}\left(m\right)$ in the chiral
limit, in agreement with equation (\ref{diverges3}). Hence the numerical results of
\cite{catorce} and \cite{sharma} are in disagreement and do not allow to get a definite
answer. Any further clarification of the numerical results for
$\chi_{\bar\pi}\left(m\right) - \chi_{\bar\delta}\left(m\right)$
would be therefore very welcome.

We have also discussed that the previous results for the two-flavor model apply also to
$N_f\ge 3$. However, universality and
renormalization-group arguments, based on the most general Landau-Ginzburg-Wilson
$\Phi^4$ theory compatible with the given symmetry-breaking pattern, make this scenario
not plausible because no stable fixed points
are expected in the corresponding Landau-Ginzburg-Wilson $\Phi^4$ theory
for $N_f\ge 3$ \cite{ettore}.


\vfill
\eject

\end{document}